\documentclass[%
reprint,
 amsmath,amssymb,
 aps,pre,
]{revtex4-1}

\usepackage{graphicx}
\usepackage{dcolumn}
\usepackage{bm}
\usepackage{xcolor}
\usepackage{hyperref}

\newcommand{\bp}{\mathrm{bp}}
\newcommand{\D}{\mathrm{d}}

\begin{document}
\title{Bifractal nature of chromosome contact maps}

\author{Simone Pigolotti}
\affiliation{Biological Complexity Unit, Okinawa Institute of Science and Technology Graduate University, Onna, Okinawa 904-0495, Japan}
\email{simone.pigolotti@oist.jp}
\author{Mogens H. Jensen}
\affiliation{The Niels Bohr Institute, Blegdamsvej 17, 2100 Copenhagen \O, Denmark}
\author{Yinxiu Zhan}
\affiliation{Friedrich Miescher Institute for Biomedical Research, Maulbeerstrasse 66, 4058 Basel, Switzerland}
\author{Guido Tiana}
\affiliation{Department of Physics and Center for Complexity and Biosystems, Universit\`a degli Studi di Milano and INFN, via Celoria 16, 20133 Milano, Italy}
\bigskip

\date{\today}

\begin{abstract}
Modern biological techniques such as Hi--C permit to measure probabilities that different chromosomal regions are close in space. These probabilities can be visualised as matrices called contact maps. In this paper, we introduce a multifractal analysis of chromosomal contact maps.  Our analysis reveals that Hi--C maps are bifractal, i.e. complex geometrical objects characterized by two distinct fractal dimensions. To rationalize this observation, we introduce a model that describes chromosomes as a hierarchical set of nested domains and we solve it exactly. The predicted multifractal spectrum is in excellent quantitative agreement with experimental data. Moreover, we show that our theory yields to a more robust estimation of the scaling exponent of the contact probability than existing methods. By applying this method to experimental data, we detect subtle conformational changes among chromosomes during differentiation of human stem cells.
\end{abstract}

\maketitle

\section{Introduction}

During cellular interphase, mammalian chromosomes assume a globular structure in the nucleus~\cite{giorgetti2014predictive,dekker2014two,tiana2016structural,dekker20163d,mccord2020chromosome}. Their conformational properties can be studied {\it in vivo} with a set of  techniques called chromosome conformation capture (3C)~\cite{dekker2002capturing}, most notably their genome-wide version called Hi--C~\cite{lieberman2009comprehensive}. Results of Hi–C experiments can be represented as matrices whose elements are proportional to the probability that two chromosomal regions are in contact in space~\cite{redolfi2019damc}. Hi--C measurements have paved the way for a mechanistic understanding of chromosome folding~\cite{tanay2013chromosomal,bonev2016organization,rowley2018organizational,szabo2019principles}. In particular, they have revealed that mammalian chromosomes are characterized by a hierarchy of nested, tightly connected structures \cite{fraser2015hierarchical,zhan2017reciprocal}. At the scale of tens of kilobases one identifies 'contact domains' \cite{sanborn2015chromatin, rao20143d}. Structures at the hundreds kilobases scale are usually called 'topological associating domains' (TADs) \cite{dixon2012topological,nora2012spatial,sexton2012three}. TADs have been extensively studied due to their essential role in gene regulation \cite{lupianez2015disruptions,symmons2016shh,flavahan2016insulator,hnisz2016activation,hanssen2017tissue,szabo2019principles}, although they do not seem to be privileged over other levels in the hierarchy from a structural point of view \cite{zhan2017reciprocal}.  At scales of few to tens of megabases, Hi--C experiments have identified 'compartments'~\cite{lieberman2009comprehensive}. Compartments are checkboard--like domains that are thought to be driven by mutually exclusive association between active and inactive chromatin \cite{lieberman2009comprehensive,falk2019heterochromatin}. At the scale of the whole nucleus, microscopy and Hi--C experiments showed that chromosomes occupy distinct 'chromosomal territories' \cite{Cremer2010, lieberman2009comprehensive}.

A simpler approach to characterize the behavior of Hi--C matrices at different scales is to study the average contact probability $p(\ell)$ of pairs of chromosomal regions $i$ and $j$ with respect to their genomic distance $\ell= |i-j|$. Such decay appears to follow a power law
\begin{equation}
p(\ell)\sim \ell^{-\beta} .
\label{eq:pij}
\end{equation}
The contact probability exponent $\beta$ is often estimated to be slightly smaller than one \cite{lieberman2009comprehensive,sanborn2015chromatin,bonev2017multiscale,zhan2017modelling}. Such low values are incompatible with simple equilibrium homopolymeric models \cite{mirny2011fractal}. In contrast, non-equilibrium models such as the crumpled globule \cite{lieberman2009comprehensive,mirny2011fractal} are able to account for such low exponents. Other proposed mechanisms include mediation of polymer interaction by other molecules \cite{barbieri2012complexity}, active loop-extrusion  \cite{fudenberg2016formation}, and finite--size effects in heteropolymers \cite{zhan2016looping}. In any case, the apparent power-law range of the contact probability is usually limited to one decade or less, see e.g. \cite{lieberman2009comprehensive,sanborn2015chromatin,bonev2017multiscale}.  Therefore, estimates of the exponent $\beta$ are rather sensitive to the choice of the fitting range.

A single physical mechanism is unlikely to explain the structure of chromatin at all scales. In fact, selective removal of proteins known to be involved in protein architecture can affect some levels of the hierarchy and not others \cite{nora2017targeted,Schwarzer2017,Haarhuis2017,Rao2017a}. Because of their importance in the control of gene expression, we focus our attention on the scales associated with TADs, i.e. below the megabase scale. It has been suggested that this hierarchy of structure can be analyzed by comparing statistical properties of Hi--C matrices at different resolutions.  \cite{chiariello2018scaling}.

In this paper, we robustly characterize scale-invariant properties of chromosomes at the scale of TADs using the theory of multifractals. This theory has been developed to characterize heterogeneous systems characterized by scale invariance  \cite{parisi1985turbulence,benzi1984multifractal,halsey1986fractal}. This analysis reveals that chromosome contact maps are {\em bifractal}, i.e. geometric objects characterized by two distinct fractal dimensions. Bifractal behavior has been previously reported in studies of surface roughness \cite{bhushan1992elastic}, distribution of matter in the universe \cite{balian1989scale} and turbulence \cite{iyer2019circulation}. We show that multifractal theory also provides a robust estimation of the scaling of the contact probability at the scale of TADs. 

The manuscript is organized as follows. In Section~\ref{sec:multifractal} we introduce the multifractal analysis using as example a Hi--C map of chromosome 1 of mouse embryonic stem cells. In Section~\ref{sec:model} we introduce the hierarchical domain model, compute its multifractal spectrum and the scaling of its contact probability. We also show that the model predicts very accurately the multifractal spectrum of Hi--C maps. In Section~\ref{sec:experiments}, we apply our findings to a broader range of experimental datasets. We show how our theory can be used as a computational method to discern differences among Hi--C maps in different experiments. We show in particular that our method is able to characterize differentiation of human stem cells. Section~\ref{sec:discussion} is devoted to conclusions and perspectives.

\section{Multifractal analysis}\label{sec:multifractal}

\begin{figure*}[t]
\begin{center}
\includegraphics[width=0.8\textwidth]{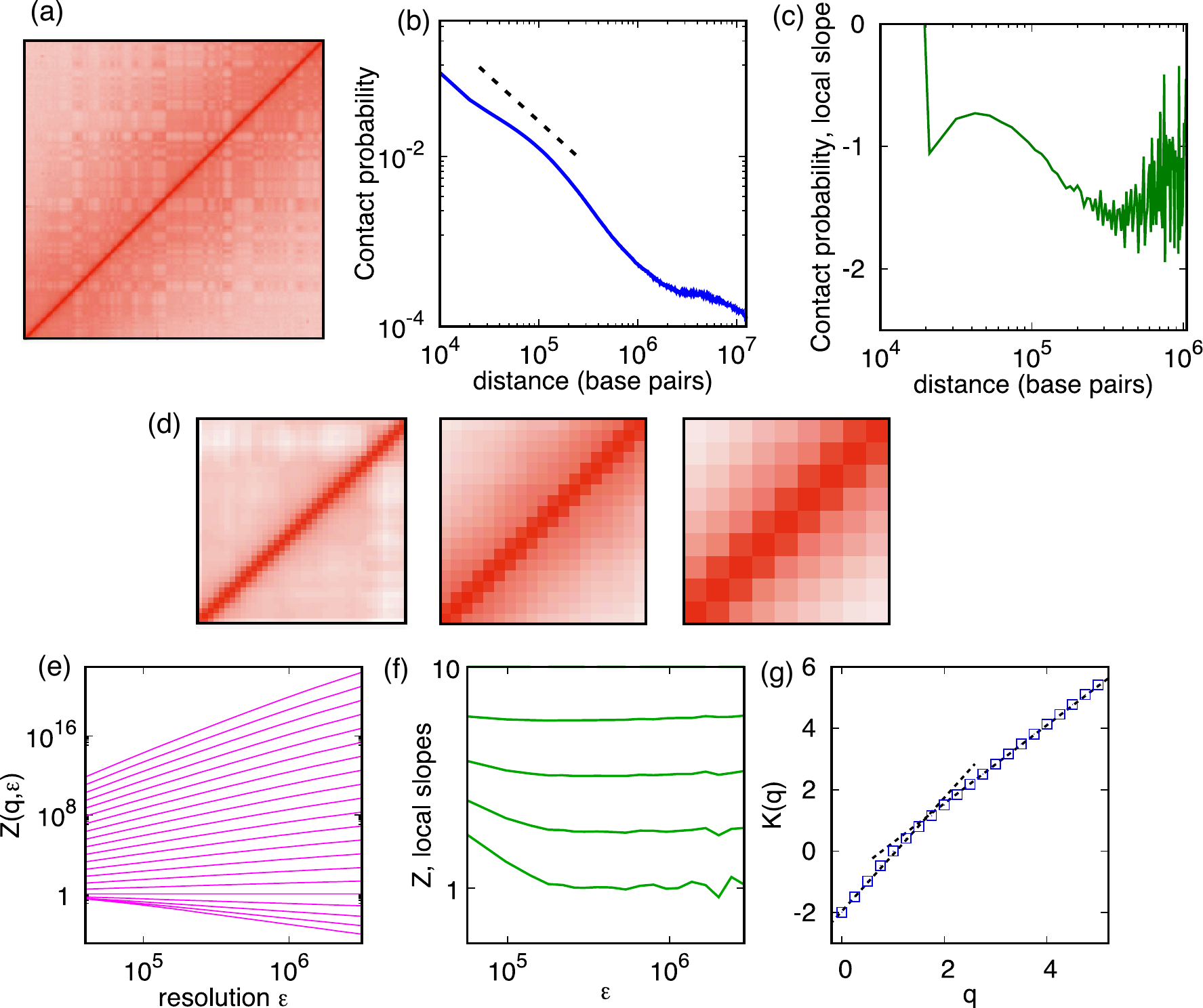}
\caption{{\bf Contact probability scaling versus multifractal analysis of a Hi--C map.} (a) Hi--C map of chromosome 1 of mouse embryonic stem cells (mESC) \cite{giorgetti2016structural}. Darker shades of red correspond to higher values of the contact probability $p_{ij}$ (b) Scaling of the average contact probability $p(\ell)$ as a function of the distance $\ell= |i-j|$. The continuous line is a best-fit of a power law, Eq. \eqref{eq:pij}, in the range of distances $[10^5,10^6]$,  yielding $\beta\approx 1.06 $. (c) Local logarithmic slope $\D \ln P(\ell)/\D (\ln \ell)$ of the contact probability. (d) Coarse-graining of the Hi--C map at increasing values of the resolution $\epsilon=10^6,10^7,2~10^7~\bp$.  (e) Scaling of $Z(q,\epsilon)$ as a function of the resolution $\epsilon$, see Eqs. \protect\eqref{eq:Z} and \protect\eqref{eq:multif}. Different lines correspond to different values of $q$, increasing from top to bottom from $q=0$ to $q=5$ at intervals of $\Delta q=0.25$. (f) Local logarithmic slopes of the first momenta of $Z(q)$ ($q=0, 0.25, 0.5, 0.75$).  (g) Corresponding multifractal spectrum $K(q)$. Here and throughout the paper, spectra are obtained by a fit in log-log scale of  $Z(q,\epsilon)$ versus $\epsilon$ in the range $\epsilon\in[10^5,10^7]$ unless specified otherwise. Two straight dashed lines are shown to guide the eye. Python code to compute the multifractal spectrum is available at \cite{pigo_code}.\label{fig:data}}
\end{center}
\end{figure*}

We introduce our idea using a Hi--C map of chromosome 1 in mouse embryonic stem cells, see Fig.~\ref{fig:data}a and Appendix~\ref{app:data}. The contact probability seems to decay as a power law, at least for relatively short genomic distances, see Fig.~\ref{fig:data}b. However, the local logarithmic slope of the contact probability does not present the clear plateau characteristic of a power law, see Fig.~\ref{fig:data}c.

To characterize scaling properties of chromosomes in a more robust way, we study the Hi--C map as a multifractal. A multifractal is a system described in terms of a density, that in our case is the density of counts in the Hi--C map. To study structures at different scales, we construct two-dimensional histograms of the Hi--C map with bins of different linear resolution $\epsilon$, see Fig.~\ref{fig:data}d. Geometrical structures of linear size smaller than $\epsilon$ are not resolved in these maps. The smallest possible value of $\epsilon$ is the resolution of the original Hi--C map, in our case $\epsilon=4\cdot 10^4 ~\bp$. We define the probability $p_{ij}(\epsilon)$ in bin at coordinates $i$, $j$ at resolution $\epsilon$. We always work with normalized maps, so that $\sum_{ij} p_{ij}(\epsilon)=1$ for all choices of $\epsilon$.

We assume the density to be scale invariant, at least for $\epsilon$ small enough:
\begin{equation}
p_{ij}(\epsilon)\sim \epsilon^{\alpha },
\end{equation}
where $\sim$ denotes the leading behavior and $\alpha$ is the scaling exponent associated with the density. Since the map is not homogeneous, different bins can be in principle  characterized by different values of $\alpha$. The number of bins $N(\alpha)$ associated with a given value of $\alpha$ must also scale as a power of $\epsilon$
\begin{equation}
N(\alpha)\sim \rho(\alpha)\epsilon^{-D(\alpha)}.
\end{equation}
The exponent $D(\alpha)$ characterizes the scaling of the number of bins of linear size $\epsilon$ necessary to cover the set with density exponent $\alpha$ and therefore can be interpreted as the fractal dimension associated with this set. The quantity $\rho(\alpha)$ is a prefactor independent of $\epsilon$.

Computing $D(\alpha)$ directly is often unpractical. A convenient related quantity is the partition function $Z(q,\epsilon)$, defined by \cite{benzi1984multifractal,halsey1986fractal}
\begin{equation}\label{eq:Z}
Z(q,\epsilon)=\sum_{ij} \left[p_{ij}(\epsilon)\right]^q .
\end{equation}
The name ``partition function'' originates from an analogy with statistical physics, where the exponent $q$ plays the role of an inverse temperature. Indeed, in an non-homogeneous system, for $q\rightarrow 0$ (high temperature), all bins give similar contributions to the sum in Eq. \eqref{eq:Z}, whereas for large $q$ (low temperature) the sum is dominated by relatively few terms characterized by largest values of the measure $p_{ij}$. For a scale-invariant system, one also expects a power-law scaling of the partition function with the resolution,
\begin{equation}\label{eq:multif}
Z(q,\epsilon) \sim \epsilon^{K(q)},
\end{equation}
at least for small $\epsilon$. This happens to be the case for our Hi--C map, see Fig.~\ref{fig:data}e. In this case, for $\epsilon$ slightly larger than its minimum value, the local logarithmic slope of the partition function is essentially flat for a broad range of scales encompassing the typical sizes of TADs, see Fig.~\ref{fig:data}f. This signals that the power--law behaviour in Eq.~\eqref{eq:multif} is much more robust than that of $p(\ell)$.

In the theory of multifractals, the function $K(q)$ defined in Eq.~\eqref{eq:multif} is called the {\em multifractal spectrum}. The multifractal spectrum is related with the fractal dimensions $D(\alpha)$ by a Legendre transform. To show that, we collect in Eq.~\eqref{eq:Z} all terms with the same value of $\alpha$:
\begin{equation}\label{eq:z2}
Z(q,\epsilon) \sim \sum_\alpha \rho(\alpha)\epsilon^{q\alpha-D(\alpha)} .
\end{equation}
A saddle-point evaluation of Eq.~\eqref{eq:z2} reveals that
\begin{equation}
K(q) = \min_\alpha [q\alpha- D(\alpha)],
\end{equation}
as anticipated. As a consequence, a linear multifractal spectrum indicates that the system is homogeneous, i.e. all of its parts are characterized by the same fractal dimension $D=\alpha$ and hence by the same scaling behavior $N\sim\epsilon^{-D}$ and $p_{ij}\sim \epsilon^{D}$. Conversely, a non-linear multifractal spectrum signals a diversity of scaling exponents and associated dimensions. 
The multifractal spectrum of chromosome 1 presents two different linear regimes, see Fig.~\ref{fig:data}g and can therefore be associated with two distinct fractal dimensions. A system with such properties is named a bifractal  \cite{bhushan1992elastic,balian1989scale,iyer2019circulation}. Since the exponent $q$ is analogous to an inverse temperature, the sharp change of slope in the spectrum of a bifractal system  is akin to a phase transition in equilibrium statistical physics \cite{bohr1987order}.

\section{Hierarchical domain model}\label{sec:model}

\begin{figure}[htb]
\begin{center}
\includegraphics[width=0.45\textwidth]{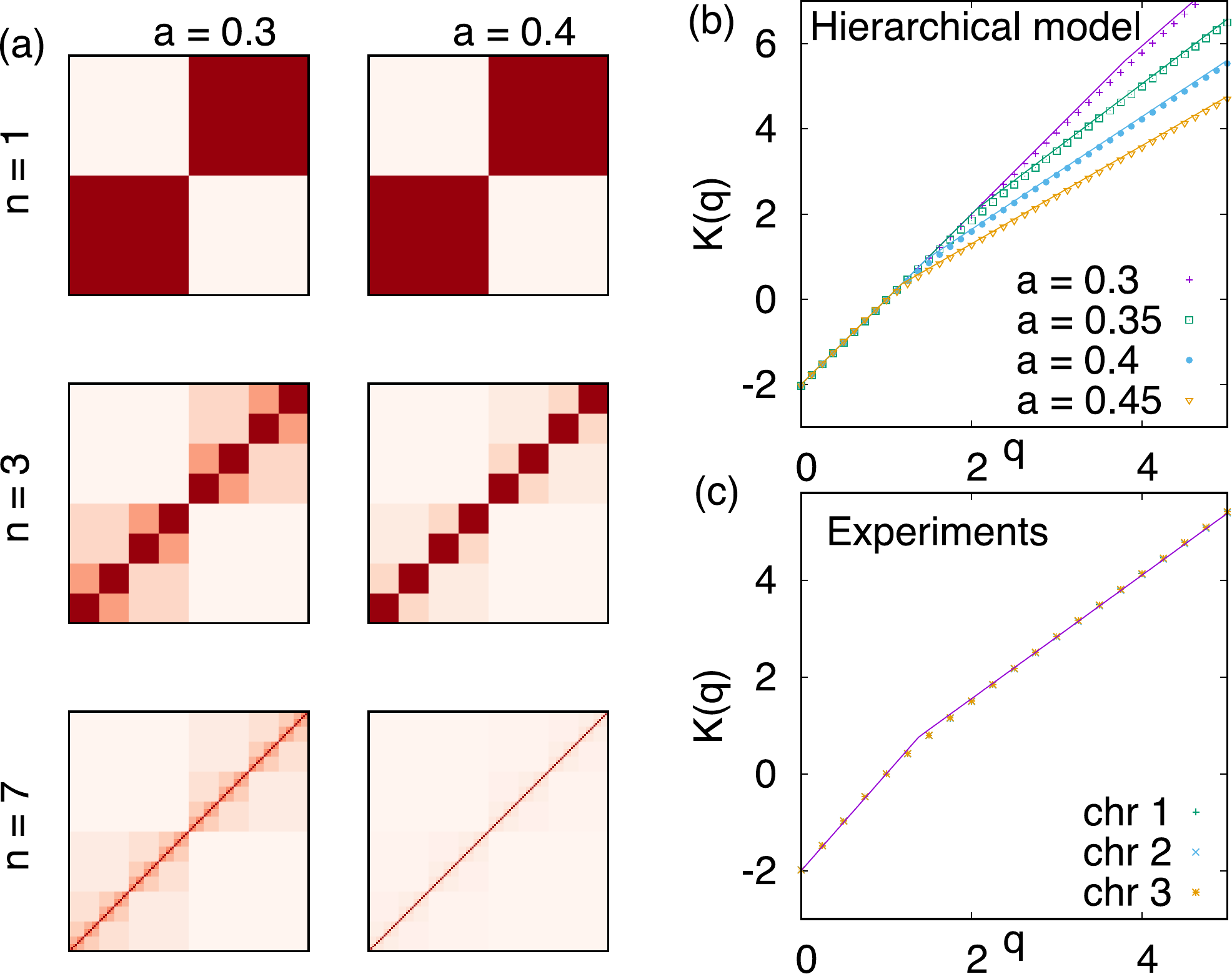}
\caption{{\bf Hierarchical domain model.} (a) Construction of the hierarchical domain model for two different values of the parameter $a$. (b) Colored symbols are the multifractal spectrum calculated by the hierarchical domain mode at different values of $a$, the solid curve is the prediction of Eqs. \eqref{eq:lowt} and \eqref{eq:hight}. (c)  Multifractal spectra of the first three chromosomes of mESC (colored symbols). The solid curve is a fit of  Eqs. \eqref{eq:lowt} and \eqref{eq:hight}, resulting in $a\approx 0.413$. Here and in the following, all fits are performed using the least square method, unless specified otherwise.
\label{fig:spectra}}
\end{center}
\end{figure}

To rationalize these observations, we introduce a hierarchical domain model of Hi--C maps.  We define the model by an iterative transformation of a measure on the unit square $[0,1]\times[0,1]$. At the first iteration, the measure is given by a $2\times 2$ matrix with diagonal element $a$ and off--diagonal elements $b$, with $a>b>0$. At each following iteration $n$ we generate a $2^n\times 2^n$ matrix. The off-diagonal element of the matrix at the previous iteration becomes a $2 \times 2$ block of identical values in the matrix at the $n$-th iteration. The diagonal blocks are further multiplied by the original matrix. The procedure is illustrated in Fig.~\ref{fig:spectra}a and Supplementary Fig. S1. We impose $b=1/2-a$, so that the measure remains normalized at each iteration. This means that, effectively, the model is defined in terms of a single free parameter $a$. Because of the normalization and the requirement that the measure should be concentrated along the diagonal, such parameter falls in the range $1/4\le a\le 1/2$.

Physically, the parameter $a$ represents the weight of domains compared to the rest of the Hi--C map. Matrices with larger $a$ have a measure more concentrated along the diagonal, whereas matrices with smaller $a$ are characterized by a more uniform measure, see Fig.~\ref{fig:spectra}a. In the limiting case $a=1/4$ the matrix is uniform at all iterations, whereas for $a=1/2$ the matrix is characterized by a uniform measure on the diagonal and all the off-diagonal elements are zero.

We now analytically compute the multifractal spectrum of the hierarchical domain model.  The partition function can be expressed as
\begin{equation}\label{eq:Zsum}
Z(q,n)=\sum_{k=0}^n \exp[\xi(k)],
\end{equation}
where we defined the exponent
\begin{eqnarray}\label{eq:Zsum2}
\xi(k)&=&[n+\max(n-k-1,0)]\ln 2+kq\ln
  a+\nonumber\\
  &+&(1-\delta_{kn})q\ln b-\max(n-k-1,0)q\ln4 .
\end{eqnarray}
A saddle--point estimation of the partition function gives
\begin{equation}
\frac{d\xi}{d k}=-\ln2+q\ln 4a,
\end{equation}
which is positive for $q>q_c=(\ln 2)/(\ln 4a)$. This means that, for $q\ge q_c$, the leading term is either $\xi(n)$ or $\xi(n-1)$. Since $\xi(n)-\xi(n-1)=q\ln(a/b)$, the maximum of the exponent is attained at $k=n$. Thus, for large $n$, the partition function scales as $Z(q,n)\sim \epsilon^{K(q)}$ with the length scale $\epsilon=2^{-n}$ and the multifractal spectrum
\begin{equation}\label{eq:lowt}
K(q)=-q \frac{\ln a}{\ln 2}-1 .
\end{equation}

For $a=1/2$, the matrix $p_{ij}(\epsilon)$ is diagonal at each iteration. In this case, Eq.~\eqref{eq:lowt} predicts a linear spectrum with slope $D=1$, consistent with the fact that the geometric set is equivalent to a one dimensional line. For $a=1/4$ the distribution is uniform on the square, and Eq.~\eqref{eq:lowt} correctly returns $D=2$. Between these two limiting cases, Eq.~\eqref{eq:lowt} predicts a fractal distribution with a dimension $D=-\ln a/\ln 2$ between $1$ and $2$.

We now focus on the ``high temperature phase'' where $-\ln2+q\ln 4a<0$. In this case, the term dominating the scaling is $k=0$, so that
\begin{equation}\label{eq:hight}
K(q)=2(q-1).
\end{equation}
Since in the high temperature phase the scaling is determined by the terms far from the diagonal, the spectrum is that of a regular two-dimensional set.

Summarizing, the predicted multifractal spectrum is characterized by two linear regimes: one with slope $-\log(a)/\log(2)$ for $q>q_c=(\ln 2)/(\ln 4a)$, Eq.~\eqref{eq:lowt}, and one with slope $2$ for $q<q_c$, Eq.~\eqref{eq:hight}.  Such predictions are in excellent agreement with numerical simulations, as shown in Fig.~\ref{fig:spectra}b, with very small discrepancies for high values of $q$ and low values of $a$ arising from finite size effects. These results confirm the validity of our saddle point approximation.

Strikingly, our theory predicted extremely well also the multifractal spectra of real chromosomes, with a fitted value of $a\approx 0.425$ and very little variability among the first three chromosomes of mouse embryonic stem cells (mean squared displacement MSD $\approx 0.0015$ in all three cases), see Fig.~\ref{fig:spectra}c. To test whether this result is a unique signature of a hierarchical mechanism, we  numerically computed the spectra of two null models. In the first null model, the contact probabilities are expressed as 
\begin{equation}\label{eq:null1}
p_{ij}=|i-j|^{-\beta}+b_{ij},
\end{equation}
where $b_{ij}$ is equal to $1$ if $i$ and $j$ belong to the same block of linear size $M$ and zero otherwise. In the second null model, the contact probabilities are given by 
\begin{equation}\label{eq:null2}
p_{ij}=|i-j|^{-\beta}+\xi_{ij}. 
\end{equation}
The terms $\xi_{ij}$ are independent, identically distributed random variables with average $0$ and standard deviation $\sigma$. If the resulting value of $p_{ij}$ is negative, it is rounded up to zero. In both models, the contact probabilities are normalized at the end of the procedure.

\begin{figure}[htb]
\begin{center}
\includegraphics[width=0.45\textwidth]{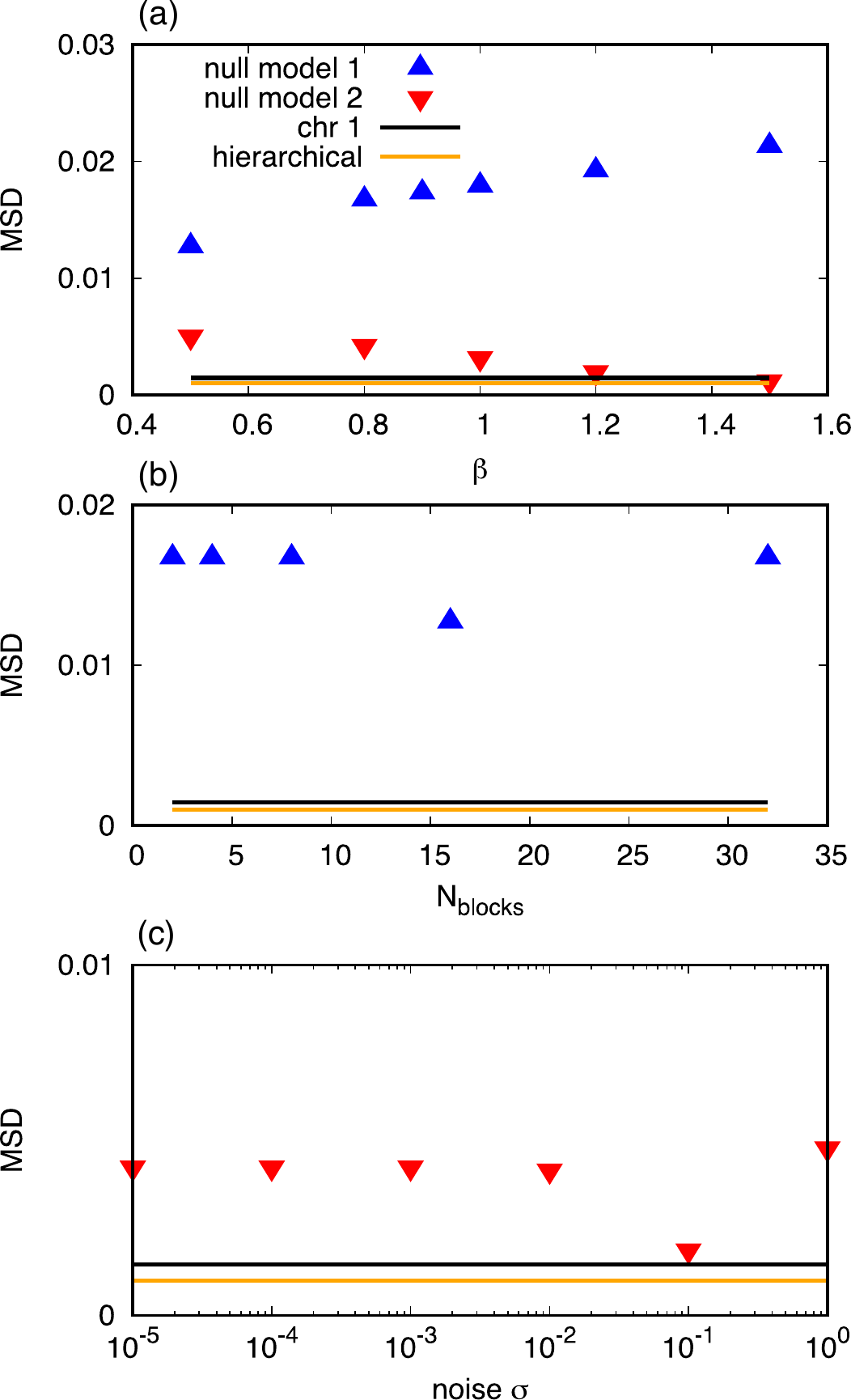}
\caption{{\bf Performance of null models.} Points and curves show the mean square deviation (MSD) between the multifractal spectrum computed from different sets of data and the one predicted by Eqs.~\eqref{eq:lowt} and \eqref{eq:hight} upon fitting the parameter $a$.
Blue upward and red downward triangles denote fits of the multifractal spectrum computed from the null models expressed by Eqs.~\eqref{eq:null1} and \eqref{eq:null2}, respectively. In (a), fits are shown as a function of $\beta$ for fixed $N_{\mathrm{blocks}}=N/M=15$ for the first null model, and fixed  $\sigma=3\ 10^{-5}$ for the second null model. In (b), fits of the first null model are shown as a function of $N_{\mathrm{blocks}}$ for fixed $\beta=0.5$. In (c), fits of the second null model are shown as a function of $\sigma$ for fixed $\beta=0.8$. In all panels, the dark black line indicates the MSD of a fit of experimental data from mouse chromosome 1. The light orange line indicates the MSD associated with a numerical fit of the hierarchical model itself, with $a=0.45$ and $n=10$ iterations. 
\label{fig:negative}}
\end{center}
\end{figure}

The solution of Eqs.~\eqref{eq:lowt} and \eqref{eq:hight} provides a poor fit to the first null model (MSD in the range $0.01 \sim 0.02$ depending on parameters), see Fig.~\ref{fig:negative}a and \ref{fig:negative}b. 
The second null model provides a better fit, comparable with that of chromosomes for some values of the parameters, see Fig.~\ref{fig:negative}. However, for realistic values of parameters ($\sigma\approx 3\ 10^{-5}$ and and $\beta\lesssim 1$) the quality of the fit is appreciably worse than that of maps from a real chromosome, see Fig.~\ref{fig:negative}a and \ref{fig:negative}c).

Our theory accurately describes also Hi--C maps of Drosophila chromosomes (average MSD $\approx 0.001$ for the first four chromosomes, see Supplementary Fig. S2) and of human chromosomes (average MSD $\approx 0.005$ for all chromosomes except chromosome Y, see Supplementary Fig. S3). These observations support that the bifractal spectrum predicted by the hierarchical domain model is compatible with that observed in a broad class of higher organisms.

\begin{figure}[htb]
\begin{center}
\includegraphics[width=0.9\linewidth]{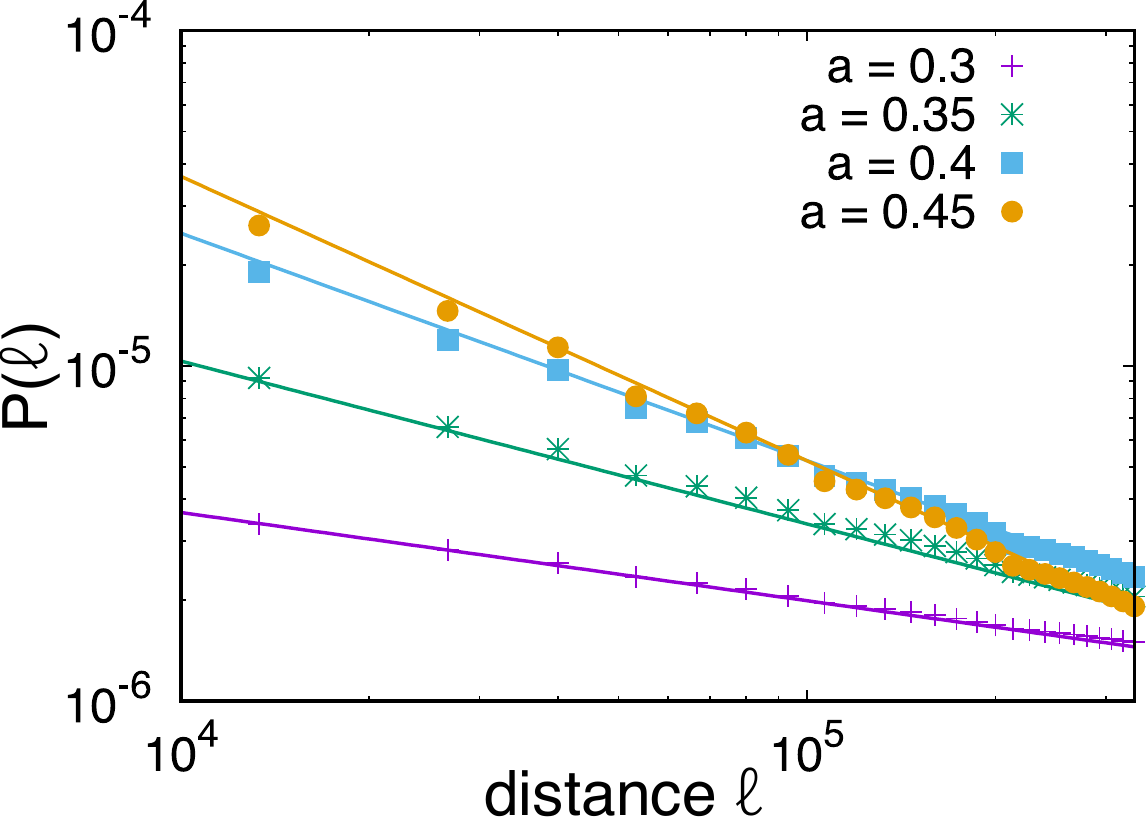}
\caption{{\bf Contact probability in the hierarchical domain model.} Decay of contact probability $P(\ell;n)$ as a function of the genomic distance $\ell$ in the hierarchical model for $n=10$ and different values of the parameter $a$, shown in the figure legend.  Continuous lines are the exponent predictions of Eq.~\eqref{eq:beta}. \label{fig:betamodel}}
\end{center}
\end{figure}

In the hierarchical domain model, the contact probability $P(\ell;n)$ decays as a power law of the genomic distance (see Eq.~\eqref{eq:pij}), with an exponent
\begin{equation}\label{eq:beta}
\beta=\frac{\ln(4a)}{\ln(2)}.
\end{equation}
Derivation of Eq.~\eqref{eq:beta} is presented in Appendix~\ref{app:scaling}. For $1/2\le a\le 1$, Eq.~\eqref{eq:beta} predicts contact probability exponents in the range $\beta\in[0,1]$. This prediction is supported by numerical simulations, see Fig.~\ref{fig:betamodel}. This means that the hierarchical structure of contact matrices automatically leads to exponents $\beta\le 1$, as observed in chromosomes.
\begin{figure*}[htb]
\begin{center}
\includegraphics[width=\linewidth]{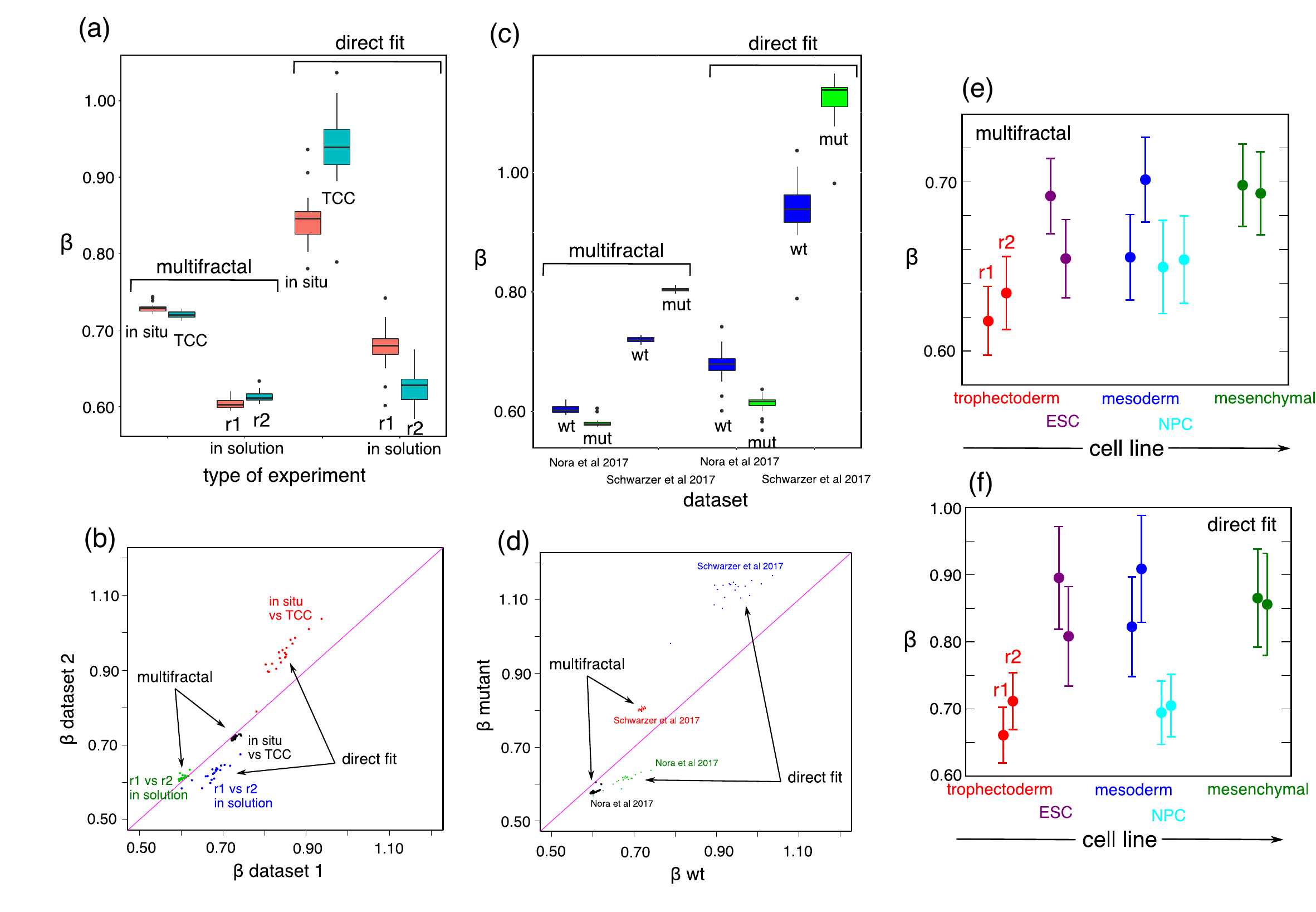}
\caption{{\bf Application of multifractal analysis to experimental Hi--C maps.} 
(a) Exponents $\beta$ calculated from the multifractal spectrum using Eq.~\eqref{eq:beta} and from a direct fit of mouse embryonic stem cells (as in Fig. \ref{fig:data}b) measured with different Hi--C methods. In particular, we analyze results of Hi--C experiments of in situ \cite{redolfi2019damc}, TCC \cite{schwarzer2017two} and two independent replicates of in solution \cite{giorgetti2016structural,nora2017targeted}. Black bars indicate  medians over chromosomes, colored boxes mark the first and third quartiles and black dots are outliers. (b) Scatter plot of the data displayed in panel a. (c) Value of $\beta$ obtained for the wild--type (wt) system and for a mutant (mut) in which either CTCF \cite{nora2017targeted} or Nipbl \cite{schwarzer2017two} is depleted in experiments.  (c)  (d) Scatter plot of the data displayed in panel c. (e) Exponents $\beta$ obtained with the multifractal method for cell lines at different stages of human stem cell development  \cite{dixon2015chromatin}. (f) Exponents obtained by direct fit of the same data.}
\label{fig:beta_alt}
\end{center}
\end{figure*}

\section{Comparison with experimental data}\label{sec:experiments}

We test more extensively our method using datasets from different Hi--C experiments.  We fit the multifractal spectrum of each dataset and obtain the corresponding exponent $\beta$ via Eq.~\eqref{eq:beta}.  Python scripts to perform this analysis on any Hi--C dataset are freely available \cite{pigo_code}. We first test the robustness of the multifractal analysis across two Hi--C experiments in mouse embryonic stem cells (mESC) from two different labs \cite{giorgetti2016structural,nora2017targeted}, both following the Hi--C protocol {\it in solution}, see Fig.~\ref{fig:beta_alt}a.  Both the multifractal and the power law methods predict that the variability of the exponent $\beta$ across chromosomes is significantly larger than the variability of the exponent across replicates. To determine which method performs best, we implement a bootstrap test with the null hypothesis that the values of $\beta$ of different chromosomes are paired at random. The multifractal method permits to exclude random pairing of chromosomes (p--value $5.2\cdot 10^{-4}$) much better than the direct fit (p--value $0.11$), see Supplementary Table S1. Moreover, the mean squared difference of $\beta$ between the two replicates is smaller in the multi--fractal case compared to direct fit ($0.009$ vs $0.04$).  The $\chi^2$ associated with the direct fit is affected by a strong systematic error, although remaining quite correlated. This effect is much milder in the multifractal approach, see Fig.~\ref{fig:beta_alt}b. The multifractal and direct fit methods are similarly robust with respect to varying the resolution of Hi--C maps, see Supplementary Fig. S4.

We apply the two methods to attempt to distinguish among experiments on the same cell lines, but following different experimental protocols. These different protocols mainly differ in the ligation step of digested DNA fragments. In the original ``in solution'' protocol, the ligation is performed in a diluted solution \cite{nagano2015comparison}. In other protocols, the ligation step is either carried out in intact nuclei (``in situ'' protocol \cite{nagano2015comparison}) or on solid substrates (``TCC'' protocol \cite{kalhor2012genome}). The two latter protocols are able to produce Hi-C maps with better signal-to-noise ratio compared to the in solution protocol \cite{nagano2015comparison,kalhor2012genome}. Both the multifractal and the direct fit methods show that the values of $\beta$ obtained from in situ and TCC experiments are markedly different from those obtained in solution, see Fig.~\ref{fig:beta_alt}a). The multifractal method estimates compatible scaling exponents for in situ and TCC protocols. The p--value associated with the null hypothesis that chromosomes are paired at random is $0.021$, see Supplementary Table S2 and Fig.~\ref{fig:beta_alt}b. A direct fit does not permit to draw this conclusion (p--value $0.14$). This result suggests that the in situ and TCC protocols result in statistically compatible Hi--C maps due to their high signal-to-noise ratio.

We compare the two methods in detecting differences between wild--type and mutant cell lines, in which either the Nipbl gene \cite{schwarzer2017two} or the CTCF gene \cite{nora2017targeted} is knocked down (see Fig.~\ref{fig:beta_alt}c). Knock--down of these genes has been shown to disrupt chromosome folding. In particular, Nipbl knock--down leads to loss of TAD structures and global changes in scaling properties \cite{schwarzer2017two}. We quantify average differences in exponents between the wild--type and the mutant in terms of $\chi^2$ of pairs of chromosomes, weighted by their mean squared errors, see Supplementary Fig. S5. Multifractal analysis detects more marked differences ($\chi^2=214$) compared to the direct fit ($\chi^2= 4.7$), although both methods highlight a statistically significant difference between the two sets (p--values $1.3\cdot 10^{-5}$ for the direct fit and $<10^{-6}$ for the multifractal). Knock--down of CTCF also causes a loss of TAD structure, but without a clear effect on genome--wide scaling \cite{nora2017targeted}. Nonetheless, the values of $\beta$ of different chromosomes obtained by multifractal analysis reveal a large ($\chi^2=86$) and significant (p--value $<10^{-6}$) difference between the scaling properties of wild--type and CTCF--deficient cells. Also in this case, the direct fit detects less marked differences ($\chi^2=25$), see Fig.~\ref{fig:beta_alt}c and Fig.~\ref{fig:beta_alt}d).

We apply the multifractal analysis to elucidate how chromosome structure changes during cellular differentiation. To this aim, we analyze Hi--C data obtained from different human cell lines at different stages of early development \cite{dixon2015chromatin}.  In temporal order, one differentiation branch includes the cell lines:  trophectoderm, embyonic stem cells (ESC), mesoderm, and mesenchymal. Another differentiation branch is in order trophectoderm, embyonic stem cells (ESC),  neuronal precursor cells (NPC) cells.
The value of $\beta$ obtained by multifractal analysis (see Fig.  ~\ref{fig:beta_alt}e) tends to increase upon differentiation. We first tested for significance applying Kendall's tau test to all chromosome pairs in different cell lines, excluding Mesoderm-NPC pairs since they are at a similar differentiation stages. We obtained a p--value  $<10^{-15}$ for the multifractal test versus $10^{-3}$ for the power law fit, see Fig.  ~\ref{fig:beta_alt}f. Testing for ordering of cell lines in each differentiation branch separately results in p--values $<10^{-18}$ (multifractal) versus $2\cdot 10^{-6}$ (power law fit) for the first branch, and $10^{-4}$ (multifractal) versus $0.5$ (power law fit) for the second branch. In this latter case, the direct power law fit is not sensitive enough to detect a significant ordering. The increase of $\beta$ with the developmental stage points to the idea that more differentiated cell types require more insulated domains to achieve more specialised function.

\section{Conclusions}\label{sec:discussion}

Multifractal analysis of Hi--C data is a powerful statistical tool to characterize scaling properties of chromosomes. We have shown that contact maps of chromosomes are bifractal, i.e. they are characterized by two distinct fractal dimensions. To understand this observation, we proposed a hierarchical domain model, whose analytical solution is in striking quantitative agreement with observed multifractal spectra. Our theory implicates a power-law scaling of the contact probability with exponents $\beta$ lower than one, in agreement with experimental observations. The multifractal method is sensitive enough to discard a null model with power-law decay of the contact probability, but domains on a single length scale only. We found that another null model, in which the contact probability is a sum of a power law plus noise, is able to produce a similar spectrum, but only for unrealistically large values of either the exponent $\beta$ or the noise intensity. The predicted form of the multifractal spectrum provides a more stringent prediction than the contact probability exponent alone. The analysis proposed here can be used as a stringent benchmark to select among different polymer models that provide similar values of $\beta$ \cite{mirny2011fractal,barbieri2012complexity,fudenberg2016formation,zhan2016looping}.

Our results indicate that scaling properties of chromosomes are a direct consequence of the hierarchical structure of chromosome domains, at least at the level of TADS \cite{rao20143d,zhan2017reciprocal}. Recent work has suggested that such domain structure is generated by a "hierarchical folding" mechanism, mediated by different proteins such as cohesin and CTCF \cite{bonev2016organization}. The precise mechanism driving the folding of chromosomes has been subject of debate \cite{rowley2018organizational}. It will be interesting to test whether the activity of these factors can produce self-similar structures compatible with our observations.  The shape of the multifractal spectrum is controlled by a single parameter, that also controls the contact probability exponent  $\beta$ via Eq. \eqref{eq:beta}. The determination of $\beta$ from the fit of the multifractal spectrum is a much more robust way of characterizing the Hi--C map than the direct fit of $\beta$. Multifractal analysis is also more sensitive in highlighting subtle structural differences as shown in the case of Nipbl and CTCF knock--down. The analysis of the multifractal spectrum is simple and robust enough to become a routinely tool to analyze contact maps, both from polymer models and experiments, and capture subtle differences among them.

\begin{appendix}

\section{Hi--C data analysis}\label{app:data}

We reanalyzed published datasets (Table 1) using HiC-Pro v.2.11.1 \cite{servant2015hic} to maintain a consistent data processing pipeline. Briefly, we mapped reads to the corresponding genome (mm10 for mouse, hg19 for human and dm6 for fly) retrieving chimeric reads and keeping only unique mappable reads. We divided genomes into bins of fixed sizes (40kb unless specified otherwise) and made an histograms of reads. We applied Ice normalization \cite{imakaev2012iterative} to the binned matrices. We then applied library size normalization to allow comparison across samples.

\begin{table*}[t]
\begin{tabular}{|l|c|c|c|c|c|}
\hline
Cell type & Organism & GEO & Condition & Protocol & Study \\
\hline

ES cells & Mus Musculus & GSE128015 & Wild Type & In situ & \cite{redolfi2019damc}\\

ES cells & Mus Musculus & GSE93431 & Wild Type & TCC & \cite{schwarzer2017two} \\

ES cells & Mus Musculus & GSE72697 & Wild Type & In solution & \cite{giorgetti2016structural} \\

ES cells & Mus Musculus & GSE98671 & Wild Type & In solution & \cite{nora2017targeted} \\

ES cells & Mus Musculus & GSE98671 & CTCF KD & In solution & \cite{nora2017targeted} \\

ES cells & Mus Musculus & GSE93431 & Nipbl KD & TCC & \cite{schwarzer2017two} \\

Trophectoderm & Homo sapiens & GSE52457 & Wild Type & In solution & \cite{dixon2015chromatin} \\

ES cells & Homo sapiens & GSE52457 & Wild Type & In solution & \cite{dixon2015chromatin}  \\

NP cells & Homo sapiens & GSE52457 & Wild Type & In solution & \cite{dixon2015chromatin}  \\

Mesoderm & Homo sapiens & GSE52457 & Wild Type & In solution & \cite{dixon2015chromatin}  \\

Mesenchymal & Homo sapiens & GSE52457 & Wild Type & In solution & \cite{dixon2015chromatin}  \\

\hline
\end{tabular}
\vspace{0.5cm}
\caption{Hi--C datasets analysed in this study. Raw data are downloaded from the GEO database at the relative accession number.}
\end{table*}

\section{Scaling of contact probability in the hierarchical domain model}\label{app:scaling}

We derive the scaling of the contact probability in the hierarchical domain model. The contact probability $P(\ell;n)$ at the $n$-th iteration can be expressed by summing the probabilities of blocks at a genomic distance $\ell$ from the diagonal:
\begin{equation}\label{eq:contprob}
P(\ell;n)=\sum_{ij}p_{ij}(n)\delta_{|i-j|,\ell}
\end{equation}
with the distance $\ell$. To find an explicit expression for $P(\ell;n)$, we start from the expression of the partition function in Eqs.~\eqref{eq:Zsum} and \eqref{eq:Zsum2}. The only difference between $P(\ell;n)$ and $Z(1,n)$ is the Kronecker delta in \eqref{eq:contprob}, that selects a subset of terms at a given genomic distance. It can be seen from the structure of the matrix that the number of terms with a given power of $a$ first increases linearly with $\ell$ up to a maximum, then decreases linearly to zero. This means that $P(\ell;n)$ can be expressed as
\begin{equation}
P(\ell;n)=\sum_{k=0}^{n-1} 2^{n} g(\ell ~ 2^{k-n}) \frac{a^k
    b^{(1-\delta_{kn})}}{4^{\max(n-k-1,0)}} ,
\end{equation}
where $g(x)$ is defined to be $x$ if $0\le x\le 1/2$ and $1-x$ if $1/2< x \le 1 $.
We now approximate the sum with an integral:
\begin{eqnarray}
P(\ell;n)&\approx& \int_{-\infty}^{n-\frac{\ln 2\ell}{\ln 2}} \D k~\ell
\exp\left[k \ln 2+k\ln a+ \right.  \\
&+&\left. (1-\delta_{kn})\ln b -\max(n-k-1,0)\ln
4\right]\nonumber\\
&+&\int_{n-\frac{\ln 2\ell}{\ln 2}}^{n-\frac{\ln \ell}{\ln 2}}  \D k~
    \exp\left[n\ln 2+ \ln \left(1-
    \ell~e^{(k-n)\ln2}\right)   \right.\nonumber\\
    &+&\left.  k\ln a+(1-\delta_{kn})\ln b -\max(n-k-1,0)\ln
4\right]\nonumber
\end{eqnarray}
For large $n$, the scaling is determined by the maximum of the argument of the exponential. This maximum is attained at a value $k^*$ given by
\begin{equation}
\ell~2^{(k^*-n)}=\frac{\ln(4a)}{\ln(4a)+\ln 2}.
\end{equation}
Therefore the decay exponent of the contact probability with the genomic distance is $\beta=\ln(4a)/\ln(2)$, as reported in Section~\ref{sec:model}.

\end{appendix}

\begin{acknowledgments}
We thank Luca Giorgetti for help with the data.
\end{acknowledgments}

\bibliography{multifractal}

\end{document}